\documentclass[11pt,onecolumn]{article}

\usepackage[utf8]{inputenc}
\usepackage[T1]{fontenc}
\usepackage{lmodern}
\usepackage{microtype}
\usepackage{geometry}
\usepackage{graphicx}
\usepackage{caption}
\usepackage{subcaption}
\usepackage{amsmath,amssymb}
\usepackage{siunitx}
\usepackage{booktabs}
\usepackage{hyperref}
\hypersetup{colorlinks=true,linkcolor=blue,citecolor=blue,urlcolor=blue}
\usepackage[sort&compress,numbers,super]{natbib}
\usepackage{authblk}
\usepackage{abstract}
\usepackage{float}
\usepackage{enumitem}
\usepackage{mathtools}
\usepackage{braket}
\usepackage{chemformula}
\usepackage{hyperref}
\usepackage{hyperref}
\hypersetup{
    colorlinks=true,
    urlcolor  = blue,
    citecolor=blue, 
    linkcolor=magenta 
}

\usepackage{orcidlink}
\usepackage{bbm}

\newcommand{\corr}{\textsuperscript{\textcolor{magenta}{\dag}}}      
\newcommand{\contri}{\textsuperscript{\textcolor{blue}{*}}} 

\begin{document}

\title{\LARGE First-principles electron-phonon scattering in real-time TDDFT}
\author[1]{Zhengwei Nie\,\orcidlink{0009-0007-1923-2660}\,\contri\corr}
\author[1]{Subhojit Pal\,\orcidlink{0009-0003-6356-3409}\,\contri\corr}
\author[2]{Martin L{\"u}ders\,\orcidlink{0000-0003-4151-4692}\,}
\author[2]{Alexander Buccheri\,\orcidlink{0000-0001-5983-8631}\,}
\author[2]{Hannes H{\"u}bener\,\orcidlink{0000-0003-0105-1427}\,}
\author[3,2]{Shunsuke A.~Sato\,\orcidlink{0000-0001-9543-2620}\,}
\author[1,2]{Umberto De Giovannini\,\orcidlink{0000-0002-4899-1304}\,\corr}
\affil[1]{Dipartimento di Fisica e Chimica-Emilio Segr\`e, \href{https://www.unipa.it/target/international-students/en/about/the-university/}{Universit\`a degli Studi di Palermo}, Via Archirafi 36, 90123 Palermo, Italy}
\affil[2]{\href{https://www.mpsd.mpg.de/person/44457/2736}{Max Planck Institute for the Structure and Dynamics of Matter}, Luruper Chaussee 149, Hamburg, Germany}
\affil[3]{Department of Physics, \href{https://www.tohoku.ac.jp/en/}{Tohoku University}, Sendai 980-8578, Japan}
\affil[ ]{\corr Corresponding author(s) E-mail(s): \href{zwnie1008@gmail.com}{zwnie1008@gmail.com},
\href{palsubhojit429@gmail.com}{palsubhojit429@gmail.com}, and \href{umberto.degiovannini@unipa.it}{umberto.degiovannini@unipa.it}}
\affil[ ]{\contri These authors contributed equally to this work.}
\date{\today}

\maketitle

\begin{abstract}
Real-time time-dependent density functional theory provides a first-principles description of coherent electron dynamics in laser-driven solids, but its unitary formulation cannot capture the irreversible scattering, relaxation, and decoherence processes that drive excited carriers toward equilibrium. Here, we develop a dissipative rt-TDDFT framework in which first-principles electron-phonon interactions enter the evolution of the reduced one-body density matrix through self-energy-derived collision integrals within the Born–Markov approximation. The approach retains the quantum-coherent real-time propagation of the electronic system while introducing phonon-mediated transitions that redistribute carriers in energy and crystal momentum, thereby incorporating the microscopic momentum-transfer processes responsible for relaxation in real materials. 
The resulting framework provides a practical first-principles route to simulate relaxation, decoherence, and time-resolved spectroscopic signatures in realistic crystalline materials.

\
\end{abstract}
\newpage

\section*{Introduction}

Real-time time-dependent density functional theory (rt-TDDFT) has become one of the most successful first-principles approaches for investigating ultrafast electron dynamics in solids driven far from equilibrium~\cite{gross1990time, marques2004time}. By explicitly propagating the electronic wavefunctions in time, rt-TDDFT provides direct access to light-induced phenomena ranging from carrier excitation~\cite{Schultze2014,Wachter2014zpt} and high-harmonic generation~\cite{Tancogne-Dejean2017,Tancogne-Dejean20177k} to Floquet engineering~\cite{Giovannini2016,Hbener2017, Neufeld2022tvm} and light-wave electronics. In parallel, the rapid development of ultrafast spectroscopies, including time- and angle-resolved photoemission spectroscopy (tr-ARPES)~\cite{Fan2025, Choi2025} and transient optical spectroscopy~\cite{Lucchini2016, Lucchini2021,Neb2026}, has enabled direct experimental observation of these nonequilibrium processes on femtosecond and attosecond timescales.

Despite these successes, a fundamental limitation remains. Conventional rt-TDDFT describes a closed quantum system and therefore generates a strictly unitary time evolution. As a consequence, it cannot account for irreversible relaxation, thermalization, and decoherence processes that arise from the coupling of excited electrons to their environment. In real materials, however, electron-phonon (e-ph) interactions provide one of the dominant channels for carrier scattering, energy dissipation, and dephasing, controlling the evolution from an optically excited nonequilibrium state back toward thermal equilibrium~\cite{giustino2017electron,bernardi2014ab}. Capturing this interplay between coherent electronic dynamics and dissipative scattering remains one of the central challenges in first-principles simulations of ultrafast materials physics.

Several approaches have been proposed to introduce dissipation into first-principles simulations, including phenomenological decoherence models and density-matrix based open-system formalisms~\cite{floss2019incorporating,oz2023electron,xu2020spin}. While these developments represent important progress, important challenges remain. In particular, a practical framework should (i) derive dissipative dynamics directly from first-principles electron-phonon interactions, (ii) consistently describe both population relaxation and coherence decay, and (iii) remain computationally viable for realistic crystalline materials.

In this work, we develop a first-principles dissipative rt-TDDFT framework in which the electronic subsystem is treated as an open quantum system weakly coupled to a thermal phonon bath. Starting from electron-phonon matrix elements obtained from first principles, we derive collision integrals for the reduced one-body density matrix within the Born-Markov approximation. The resulting formalism simultaneously captures coherent electronic dynamics, phonon-assisted carrier scattering, and the decay of quantum coherences within a unified real-time framework.

In contrast to semiclassical Boltzmann approaches, which evolve only carrier populations, the present framework retains the full single-particle density matrix and therefore provides access to both relaxation and decoherence dynamics.

A key advantage of the present implementation is that the dissipative dynamics is fully integrated within the real-space, real-time framework of the \texttt{Octopus} code~\cite{Marques2003, Castro20069rt, Andrade2012,Andrade2015,tancogne2020octopus}. The electron-phonon interaction acts only within a selected active space through the reduced density matrix, while the underlying real-time propagation retains access to the full set of observables available in real-space real-time TDDFT. This allows microscopic electron-phonon scattering processes to be connected to experimentally accessible observables using existing first-principles methodologies already available within the code. In particular, the present implementation can be naturally combined with state-of-the-art tr-ARPES simulations based on tSURFF~\cite{Giovannini2017}, providing a direct route from dissipative carrier dynamics to photoemission experiments. This combination of open-system dynamics with real-space real-time simulations provides a flexible platform for investigating nonequilibrium phenomena across a broad range of materials and experimental conditions.

This paper is organized as follows. We first present the theoretical formulation and numerical implementation of dissipative rt-TDDFT. We subsequently validate this methodology through benchmark applications to monolayer WS$_2$ and bulk Si, demonstrating its ability to capture intervalley scattering driven by e-ph coupling as well as ultrafast carrier thermalization. Lastly, we address the broader implications of the methodology and future outlook for first-principles simulations of dissipative quantum dynamics in materials.

\section*{Results}

\subsection*{Coherent real-time TDDFT dynamics}

In conventional real-time time-dependent density functional theory (rt-TDDFT), the electronic dynamics of a crystalline solid are described through the propagation of auxiliary single particle system represented by time-dependent Kohn-Sham (TDKS) Bloch orbitals, $|\varphi_{n\mathbf k}(t)\rangle$. For each band index $n$ and crystal momentum $\mathbf k$, these orbitals obey (atomic units are used throughout)
\begin{equation}
i\frac{\partial}{\partial t}|\varphi_{n\mathbf k}(t)\rangle =\hat H_{\rm KS}(t) |\varphi_{n\mathbf k}(t)\rangle .
\label{eq:tdks}
\end{equation}
In the velocity gauge, the time-dependent Kohn-Sham Hamiltonian is written schematically as
\begin{equation}
\hat H_{\rm KS}(t) =
\frac{1}{2} \left(\hat{\mathbf p} - \frac{\mathbf A(t)}{c}
\right)^2 +\hat V_{\rm ion}+\hat V_{\rm H}[n(t)]+\hat V_{\rm xc}[n(t)] .
\label{eq:hks}
\end{equation}
Here $\hat{\mathbf p}$ is the momentum operator, $\mathbf A(t)$ is the time-dependent vector potential, $\hat V_{\rm ion}$ is the ionic potential, and $\hat V_{\rm H}[n(t)]$ and $\hat V_{\rm xc}[n(t)]$ are the Hartree and exchange-correlation potentials evaluated from the time-dependent density. The electronic density is obtained from the occupied TDKS orbitals as
\begin{equation}
n(\mathbf r,t) = \sum_{n\mathbf k} f_{n\mathbf k} \left| \varphi_{n\mathbf k}(\mathbf r,t) \right|^2 ,
\label{eq:density_tddft}
\end{equation}
where $f_{n\mathbf k}$ are the orbital occupations. In the coherent closed-system propagation of conventional rt-TDDFT, these occupations are fixed and the time dependence of the density arises from the evolution of the orbitals.

Equivalently, the coherent Kohn-Sham dynamics can be written in terms of the auxiliary Kohn-Sham one-particle density operator
\begin{equation}
\hat\rho_{\rm KS}^{(1)}(t) = \sum_{n\mathbf k} f_{n\mathbf k} |\varphi_{n\mathbf k}(t)\rangle \langle\varphi_{n\mathbf k}(t)| .
\label{eq:ks_density_operator}
\end{equation}
Using Eq.~(\ref{eq:tdks}), this operator obeys the Liouville–von Neumann equation
\begin{equation}
\frac{\partial\hat\rho_{\rm KS}^{(1)}(t)}{\partial t}= - i \left[ \hat H_{\rm KS}(t), \hat\rho_{\rm KS}^{(1)}(t) \right] .
\label{eq:lvn_closed}
\end{equation}
Equation~(\ref{eq:lvn_closed}) is the Liouville–von Neumann form of the auxiliary noninteracting Kohn-Sham evolution. It should not be interpreted as a closed equation of motion for the exact interacting many-body one-particle density matrix.

The evolution generated by Eq.(\ref{eq:tdks}) or, equivalently, Eq.(\ref{eq:lvn_closed}) is unitary and preserves the occupations of the propagated Kohn-Sham orbitals. It therefore captures coherent laser-driven electronic motion, including nonlinear response and interband excitation, but does not by itself describe irreversible carrier relaxation, thermalization, or decoherence arising from coupling to lattice vibrations. These processes require an open-system extension of the Kohn-Sham density-matrix dynamics.

\subsection*{Open-system density-matrix formulation}

To incorporate relaxation and decoherence, we regard the electronic degrees of freedom as an open subsystem coupled to a phonon environment. At the many-body level, the total Hamiltonian is partitioned as
\begin{equation}
\hat H(t) = \hat H_{\rm S}(t) + \hat H_{\rm E} + \hat H_{\rm S\mbox{-}E},
\label{eq:system_bath_hamiltonian}
\end{equation}
where $\hat H_{\rm S}(t)$ describes the electronic subsystem, $\hat H_{\rm E}$ the phonon environment, and $\hat H_{\rm S\mbox{-}E}$ their coupling. In the practical formulation used here, the electronic subsystem described at the Kohn-Sham single-particle level, so that $\hat H_{\rm S}(t)$ is represented by the time-dependent Kohn-Sham Hamiltonian $\hat H_{\rm KS}(t)$ introduced in Eq.~(\ref{eq:hks}), while the phonon degrees of freedom are traced out.

The microscopic electron-phonon coupling is most conveniently written in second quantization as
\begin{equation}
\hat H_{\rm S\mbox{-}E}
= \sum_{mn\mathbf k\mathbf q\lambda} g^{\lambda}_{mn}(\mathbf k,\mathbf q)\, \hat c^{\dagger}_{m,\mathbf k+\mathbf q} \hat c_{n,\mathbf k} \left(
\hat b_{\mathbf q\lambda} + \hat b^{\dagger}_{-\mathbf q\lambda} \right),
\label{eq:Helph}
\end{equation}
where $g^{\lambda}_{mn}(\mathbf k,\mathbf q)$ are first-principles electron-phonon matrix elements, $\lambda$ labels the phonon branch, and $\hat c^{\dagger}$, $\hat c$, $\hat b^{\dagger}$, and $\hat b$ are electronic and phononic creation and annihilation operators. This second-quantized form is used to define the microscopic scattering channels and the corresponding transition rates. 

Starting from the full system-bath density matrix and tracing over the phonon degrees of freedom gives an effective equation for the reduced electronic density matrix. The general derivation follows the standard Born--Markov construction for an open quantum system and is summarized in the Supplementary Information for the case of a phonon bath in thermal equilibrium. In the present work, the electronic hierarchy is then closed at the Kohn-Sham single-particle level. We therefore propagate an effective reduced single-particle density matrix, denoted $\hat\rho^{(1)}$, whose coherent evolution is generated by $\hat H_{\rm KS}(t)$ and whose dissipative evolution is generated by an electron-phonon collision integral,
\begin{equation}
\frac{\partial \hat\rho^{(1)}(t)}{\partial t}
= -i\left[ \hat H_{\rm KS}(t), \hat\rho^{(1)}(t) \right]
+ \mathcal I_{\rm e-ph} \left[ \hat\rho^{(1)}(t) \right].
\label{eq:open_density_matrix}
\end{equation}
Here $\hat\rho^{(1)}$ should be understood as the Kohn-Sham single-particle density matrix extended to include non-unitary population and coherence dynamics. In the absence of the collision integral, it reduces to the auxiliary Kohn-Sham density operator of Eq.~(\ref{eq:ks_density_operator}). The additional term $\mathcal I_{\rm e-ph}$ encodes the phonon-mediated scattering processes that remain after the bath has been integrated out.

In a single-particle basis the Born--Markov collision integral takes the form
\begin{equation}
\begin{aligned}
    &\mathcal{I}_{\rm e-ph}[\rho^{(1)}_{mn}(\mathbf{k},t)] = - (1-\delta_{mn}) \frac{\gamma_{m}(\mathbf k, t) + \gamma_{n}(\mathbf k, t)}{2} \rho^{(1)}_{mn}(\mathbf k, t) \\
    &+ \delta_{mn}\sum_{i\mathbf k'} \Big[\overline{W}_{(n,\mathbf k)\leftarrow (i, \mathbf k')} \rho^{(1)}_{ii}(\mathbf k', t) (1-\rho^{(1)}_{nn}(\mathbf k, t)) - \overline{W}_{(i,\mathbf k')\leftarrow (n, \mathbf k)} \rho^{(1)}_{nn}(\mathbf k, t) (1-\rho^{(1)}_{ii}(\mathbf k', t)) \Big] 
\end{aligned}
\label{eq:collision_integral}
\end{equation}
where $m,n,i$ label electronic bands and $\mathbf k,\mathbf k'$ denote crystal momenta.  The resulting collision integral is the single-particle reduction of an underlying completely positive trace-preserving~\cite{simoni2025first} (CPTP) open-system evolution. The quantities $\overline{W}_{(m,\mathbf k)\leftarrow(n,\mathbf k')}$ [see Eq.~(\ref{w_bar})] are coarse-grained first-principles electron-phonon transition rates obtained from the microscopic matrix elements in Eq.~(\ref{eq:Helph}). They encode phonon-assisted transfer of energy and crystal momentum between electronic states. The state-dependent scattering rate entering the coherence decay is
\begin{equation}
\gamma_n(\mathbf k,t)
= \sum_{i,\mathbf k'} \Big[ \overline{W}_{(i,\mathbf k')\leftarrow(n,\mathbf k)} \left( 1-\rho^{(1)}_{ii}(\mathbf k',t) \right)
+\overline{W}_{(n,\mathbf k)\leftarrow(i,\mathbf k')} \rho^{(1)}_{ii}(\mathbf k',t)\Big] .
\label{eq:gamma}
\end{equation}

The structure of Eq.~(\ref{eq:collision_integral}) has the usual interpretation of a density-matrix collision integral. The diagonal part has the gain--loss form of a Boltzmann equation and describes phonon-assisted redistribution of carrier populations across bands and momenta, including Pauli blocking of the final states. The off-diagonal part describes the decay of interband coherences induced by the same microscopic scattering channels. Thus, the formulation retains the connection to first-principles electron-phonon scattering theory while extending the population-only Boltzmann picture to include coherence decay.

\subsection*{Coupling active-space dissipation to real-space propagation}

Equation~(\ref{eq:open_density_matrix}) defines the open-system structure of the theory. In the actual implementation, however, the coherent and dissipative parts of the dynamics are represented in different but complementary spaces. The coherent Kohn-Sham propagation is carried out in real space, as in standard rt-TDDFT, while the electron-phonon collision integral is applied in a selected active space. The working form of the propagation can therefore be summarized as
\begin{equation}
\frac{\partial \hat\rho^{(1)}(t)}{\partial t}
=
-i
\left[
\hat H_{\rm KS}(t),
\hat\rho^{(1)}(t)
\right]_{\rm RS}
+
\mathcal I_{\rm e-ph}
\left[
\hat\rho^{(1)}(t)
\right]_{\rm AS}.
\label{eq:open_density_matrix_RS_AS}
\end{equation}
Here the subscript RS indicates that the coherent evolution is performed by propagating the Kohn-Sham orbitals on the real-space grid, whereas the subscript AS indicates that the dissipative update is evaluated after projecting the electronic state onto an active single-particle subspace. This separation is central to the method: real-space propagation preserves the full rt-TDDFT description of nonlinear light-matter dynamics and spectroscopic observables, while the active space provides a controlled representation of the electronic states involved in phonon-mediated scattering, relaxation, and decoherence.

The active-space density matrix is constructed by projecting the propagated Kohn-Sham state onto a time-dependent single-particle basis $\{|\phi_{n\mathbf k}(t)\rangle\}$,
\begin{equation}
\hat\rho^{(1)}_{\rm AS}(\mathbf k,t)
=
\sum_{mn}
\rho^{(1)}_{mn}(\mathbf k,t)
|\phi_{m\mathbf k}(t)\rangle
\langle\phi_{n\mathbf k}(t)| .
\label{eq:active_density_matrix}
\end{equation}
In this work we choose $\{|\phi_{n\mathbf k}(t)\rangle\}$ as the adiabatic Kohn-Sham basis obtained from the instantaneous Hamiltonian $\hat H_{\rm KS}(t)$. This choice is important in the presence of strong external fields. The intraband motion driven by the vector potential is absorbed into the time-dependent adiabatic basis, so that the active-space populations describe genuine interband and intervalley carrier redistribution rather than the reversible acceleration of Bloch states. Moreover, the active-space projectors are defined from the instantaneous Hamiltonian and are therefore consistently transformed with the field-dressed Kohn-Sham states. This avoids introducing gauge-dependent projections associated with fixed nonlocal projectors during the real-time propagation.

The use of the adiabatic basis also provides a natural connection to the electron-phonon transition rates. The rates $\overline W_{(m,\mathbf k)\leftarrow(n,\mathbf k')}$ are computed from first-principles electron-phonon matrix elements between static band states. During the real-time propagation, the instantaneous adiabatic states remain continuously connected to these reference band states. This makes it possible to apply the same microscopic scattering channels during the driven dynamics while maintaining a consistent band and momentum labeling of the active-space density matrix. The detailed construction of the active basis, including the residual-space completion used to avoid loss of norm under projection, is described in the Methods.
\begin{figure}[H]
  \centering
  \includegraphics[width=0.9\linewidth]{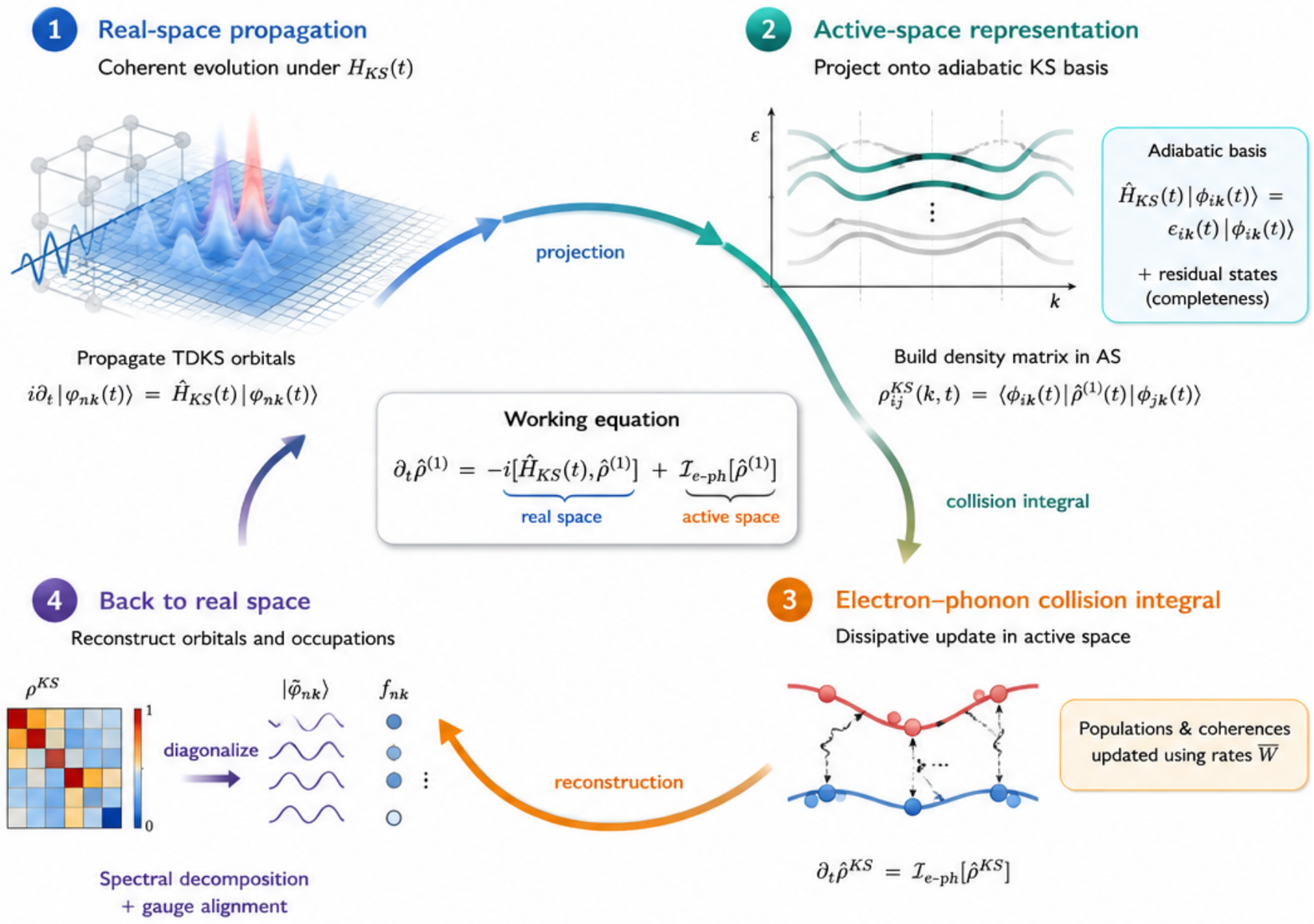}
\caption{\textbf{Real-space/active-space propagation cycle.}
Coherent rt-TDDFT dynamics are propagated in real space, while electron-phonon scattering is applied to the reduced single-particle density matrix in an adiabatic active space. After the dissipative update, the density matrix is diagonalized and gauge-aligned to reconstruct orbitals and occupations for the next real-space propagation step.}
\label{fig:rs_as_workflow}
\end{figure}

The complete workflow is summarized in Fig.~\ref{fig:rs_as_workflow}. At each time step, the Kohn-Sham orbitals are first propagated coherently in real space under $\hat H_{\rm KS}(t)$. The propagated state is then projected onto the adiabatic active basis, where the density matrix $\rho^{(1)}_{mn}(\mathbf k,t)$ is updated by the collision integral in Eq.~(\ref{eq:collision_integral}). In practice, the coherent and dissipative updates are applied sequentially within each time step; the numerical details of this splitting are given in the Methods. This procedure avoids explicit propagation of the full electron-phonon many-body problem while retaining the microscopic scattering information through the first-principles transition rates $\overline W$.

After the dissipative update, the active-space density matrix must be mapped back to a set of orbitals suitable for the next real-space TDKS propagation. We therefore recast the updated density matrix in its spectral, or natural-orbital, representation,
\begin{equation}
\hat\rho^{(1)}(t+\Delta t)
|\varphi_{n\mathbf k}(t+\Delta t)\rangle
=
f_{n\mathbf k}(t+\Delta t)
|\varphi_{n\mathbf k}(t+\Delta t)\rangle ,
\label{eq:rho_diagonalization}
\end{equation}
where $|\varphi_{n\mathbf k}(t+\Delta t)\rangle$ are the reconstructed orbitals and $f_{n\mathbf k}(t+\Delta t)$ are the corresponding nonequilibrium occupations. In this representation, the updated density matrix is equivalently described by natural orbitals with generally fractional occupations, reflecting the non-unitary open-system evolution and the associated Kohn-Sham ensemble character. The arbitrary phases and possible rotations within nearly degenerate subspaces are fixed by aligning the reconstructed orbitals to those from the previous time step, as described in the Methods. The resulting orbitals and occupations define the initial state for the next real-space propagation step.

This construction provides the practical link between open-system electron-phonon dynamics and real-space rt-TDDFT. The active-space density matrix carries the dissipative population and coherence dynamics, while the real-space propagation retains the nonlinear, continuum, and spectroscopic capabilities of the underlying rt-TDDFT framework.
We next apply this construction to two complementary regimes of electron-phonon-mediated relaxation. In bulk Si, we benchmark hot-carrier relaxation through the redistribution of photoexcited carriers in energy and momentum. In monolayer WS$_2$, we follow intervalley scattering in momentum space and connect the microscopic carrier dynamics to fully ab initio tr-ARPES observables. These applications demonstrate how dissipative rt-TDDFT links first-principles scattering mechanisms to experimentally accessible ultrafast signatures.

\subsection*{Ultrafast carrier relaxation dynamics in silicon}

We first apply the method to bulk silicon as a benchmark for electron-phonon-mediated hot-carrier relaxation in a well-characterized semiconductor. Silicon is the basis of modern optoelectronics and photovoltaics, where the redistribution and cooling of photoexcited carriers set fundamental limits to device performance~\cite{shockley2018detailed,ross1982efficiency}. It therefore provides a stringent test for the present framework: after optical excitation, the electronic system is initially driven into a highly non-equilibrium distribution, and its subsequent evolution requires phonon-mediated scattering in both energy and momentum.

\begin{figure}
  \centering
  \includegraphics[width=0.85\linewidth]{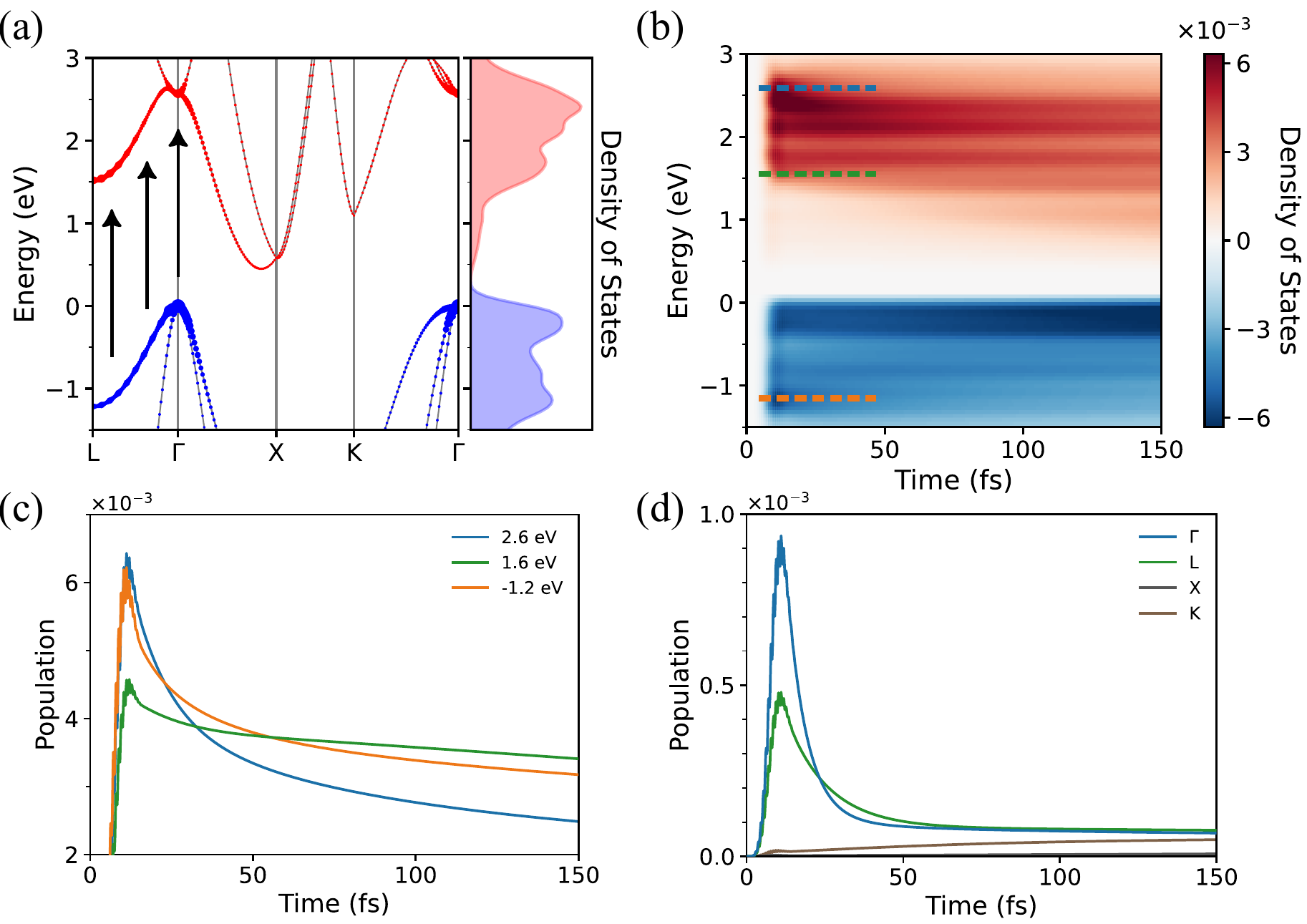}
  \caption{\textbf{Ultrafast carrier relaxation dynamics in bulk silicon.} 
  \textbf{a}, Transient band structure and projected carrier distribution upon optical excitation. Red and blue markers denote photoexcited electrons and holes, respectively. The right panel displays the corresponding occupied density of states. 
  \textbf{b}, Temporal evolution of the occupied density of states. The colored dashed lines indicate the specific energy slices selected for the population analysis in \textbf{c}. 
  \textbf{c}, Energy-resolved population dynamics. The blue, green, and orange curves correspond to temporal profiles extracted at 2.6 eV ($\Gamma$ valley), 1.6 eV (conduction-band $L$ valley), and -1.2 eV (valence-band $L$ valley), respectively. 
  \textbf{d}, Momentum-resolved temporal evolution of the excited electron populations integrated around the $\Gamma$ (blue), $L$ (green), $X$ (gray), and $K$ (brown) conduction band minima.}
  \label{fig:silicon}
\end{figure}

As illustrated in Fig.~\ref{fig:silicon}(a), resonant excitation by a 2.5 eV laser pulse promotes valence-band electrons into the conduction bands, creating non-equilibrium holes (blue) and electrons (red). Photoexcited electrons are predominantly distributed along the $\Gamma$--L path due to optical selection rules, with the corresponding occupied density of states (ODOS) shown in the right panel. In agreement with the projected band structure, holes occupy the energy range from $-1.2$ to 0 eV, while electrons span from 1.5 to 2.6 eV. The laser pulse therefore prepares a strongly non-equilibrium carrier distribution whose relaxation cannot be described within conventional unitary rt-TDDFT.

Electron-phonon coupling provides the dominant dissipation channel for the subsequent relaxation of the photoexcited carriers~\cite{bernardi2014ab, marini2013competition, fischetti1988monte, wang2024real}. Fig.~\ref{fig:silicon}(b) shows the temporal evolution of the ODOS over the first 150 fs of the relaxation process. Immediately after excitation, the populations of high-energy conduction states and deep-level valence holes deplete within tens of femtoseconds, accompanied by a spectral weight accumulation at the band edges. These phenomena signify an efficient population transfer. To gain deeper insights into the fundamental scattering mechanisms, we perform an energy-resolved population analysis of the transient ODOS. For primary valleys exhibiting population maxima upon excitation, we extract their temporal profiles as fingerprints of the evolution. Specifically, we track the temporal evolution at discrete energy slices that capture the initial populations of these key valleys: 2.6 eV (associated with the conduction-band $\Gamma$ valley, blue dashed line), 1.6 eV (conduction-band $L$ valley, green dashed line), and -1.2 eV (valence-band $L$ valley, orange dashed line).


As illustrated in Fig.~\ref{fig:silicon}(c), driven by the laser field, the carrier populations in the primary valleys undergo a rapid increase and reach a peak at about 10 fs. Subsequently, a pronounced decay is observed. To quantitatively evaluate the dissipation dynamics, we fit these curves to an exponential decay. The extracted characteristic energy lifetimes are 40 fs at 2.6 eV, 106 fs at 1.6 eV, and 45 fs at -1.2 eV. Crucially, the conduction-band energy lifetimes are in quantitative agreement with experimental measurements of 30 fs at 2.6 eV and 120 fs at 1.6 eV~\cite{tanimura2019ultrafast} (values shifted from the original CBM reference to our VBM reference for direct comparison). This quantitative consistency verifies the reliability of our dissipative dynamics approach in describing ultrafast carrier relaxation processes.


To further dissect the microscopic scattering pathways, we perform a momentum-resolved analysis targeting the photoexcited electrons. By integrating the electron density over localized momentum-space volumes around the corresponding band minima, we obtain the electron populations in the $\Gamma$ (blue), $L$ (green), $X$ (gray), and $K$ (brown) valleys, as depicted in Fig.~\ref{fig:silicon}(d). Similar to the trend observed in the energy-resolved analysis, the populations in these localized valley minima undergo a depletion following their initial peak. 
However, the characteristic momentum lifetimes extracted via exponential fitting for the $\Gamma$ and $L$ valleys are 8 fs and 14 fs, respectively. While these values approach the Markovian validity limit set by the maximum Si phonon energy ($\sim$65 meV, or $\sim$10 fs bath correlation time) and thus inevitably involve transient non-Markovian effects, this does not alter our broader conclusions on the relaxation trends. Notably, these momentum lifetimes remain much shorter than their energy-resolved counterparts of 40 fs and 106 fs.
This discrepancy indicates that electrons are scattered out of the valleys before relaxing from their specific energy ranges, implying an ultrafast momentum redistribution with minimal energy dissipation. Driven by these rapid scattering processes, a quasi-equilibrium hot-electron ensemble (HEE) is built up before the macroscopic energy decay occurs~\cite{tanimura2019ultrafast}. Additionally, we observe a gradual population accumulation in the $X$ valley, which also emerges in Fig.~\ref{fig:silicon}(b) as an increase in spectral weight at the conduction band minimum (CBM). Although this accumulation is not significant in the initial 150 fs of our simulation, it is the precursor to the subsequent cascade cooling and intra-valley thermalization, which unfold on a much longer picosecond timescale.

\subsection*{Ultrafast Valley Dynamics and Intervalley Scattering in WS$_2$}

In monolayer transition metal dichalcogenides (TMDs), the optical and transport properties are heavily dictated by their carrier relaxation mechanisms\cite{ceballos2017ultrafast,bertoni2016generation,kumar2021spin}. For example, in WS$_2$, following optical excitation, electrons thermalize rapidly due to the strong electron-phonon interactions. During this process, electrons undergo ultrafast intervalley scattering, where the population redistributes in momentum space, accompanied by the quenching of the initial valley polarization\cite{wallauer2021momentum,kolesnichenko2024sub,zhu2025holistic}. To unravel these sub-picosecond dynamics, it is crucial to track the electron relaxation upon excitation with precise time resolution.

\begin{figure}[H]
  \centering
  \includegraphics[width=0.7\linewidth]{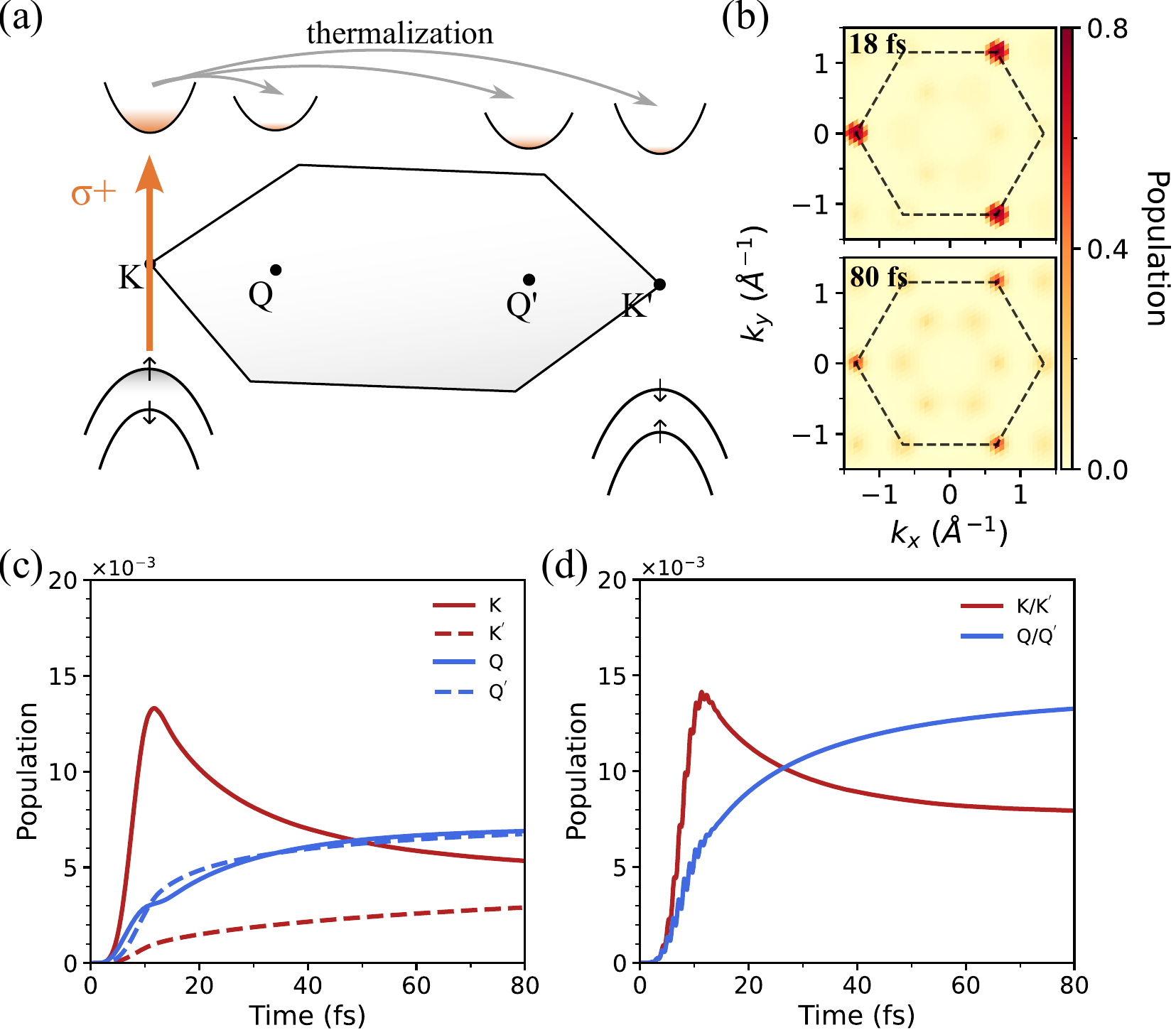}
  \caption{\textbf{Ultrafast valley scattering dynamics in monolayer WS$_{\mathbf{2}}$.} 
  \textbf{a}, Schematic of the band structure and valley-selective excitation. The spin splitting at VBM is indicated by up/down arrows. Orange and gray arrows denote the CP light excitation and subsequent phonon-mediated intervalley scattering, respectively.
  \textbf{b}, Momentum-resolved excited electron populations in the first Brillouin zone at 18 fs (upper panel) and 80 fs (lower panel), illustrating the sub-picosecond carrier transfer. 
  \textbf{c}, Temporal evolution of the valley-specific electron populations under CP light excitation. Red solid and dashed lines represent the populations in the K and K$^\prime$ valleys, respectively, while blue solid and dashed lines correspond to the Q and Q$^\prime$ valleys. 
  \textbf{d}, Temporal evolution of the valley populations under LP light excitation.}
  \label{fig:valleydynamics}
\end{figure}

Fig.~\ref{fig:valleydynamics}(a) illustrates the schematic of the band structure in the hexagonal Brillouin zone of monolayer WS$_2$. The broken spatial inversion symmetry and strong spin-orbit coupling result in the spin splitting at valence band maxima (VBM) of non-equivalent $K$ and $K^\prime$ valleys. Upon illumination by circularly polarized (CP) light, the $K$ valley is selectively populated due to the valley-contrasting optical selection rules\cite{xiao2012coupled,jones2013optical}, indicated by the orange arrow. Following this initial valley polarization, the photoexcited electrons are scattered to the $K^\prime$ valley (the counterpart of the K valley) and the Q/Q$^\prime$ valleys, which are located midway along the $\Gamma$--K and $\Gamma$--K$^\prime$ high-symmetry lines. These phonon-assisted transitions are the dominant dissipation channels for this ultrafast thermalization.

Following the excitation by a 2 eV pump pulse at 300 K, the momentum-resolved excited electron populations at 18 fs (upper panel) and 80 fs (lower panel) are shown in Fig.~\ref{fig:valleydynamics}(b). The excited electron populations are defined as the number of electrons in the conduction bands. At 18 fs, the $K$ valley is selectively populated, surpassing the excitation in $K^\prime$ and $Q$/$Q^\prime$ valleys. By contrast, at 80 fs, due to the phonon-mediated intervalley scattering, population in $K$ valley has markedly decreased, accompanied by a significant population increase in the other valleys.


The quantitative temporal evolution of the valley populations is shown in Fig.~\ref{fig:valleydynamics}(c). The dissipation can be roughly divided into a faster channel (to $Q$/$Q^\prime$ valleys) and a slower channel (to $K^\prime$ valley). For the former, population in $Q$ and $Q^\prime$ valleys differ within the initial 30 fs due to the anisotropic nature of their specific e-ph couplings, and equalize thereafter. As for the latter, the slower population increase is attributed to the weaker scattering strength between $K$ and $K^\prime$ valleys. By fitting the depletion of the $K$ valley population after pump pulse to an exponential decay, we extract a characteristic intervalley scattering time of 20 fs.


Fig.~\ref{fig:valleydynamics}(d) shows the valley population evolution under illumination by linearly polarized (LP) light. In contrast to the CP light case, the $K$ and $K^\prime$ valleys are equally excited, resulting in a symmetric population increase in $Q$/$Q^\prime$ valleys. An exponential fit to the $K$-valley population depletion again yields a characteristic lifetime of 17 fs, which is in good agreement with recent experimental value of 16 fs~\cite{wallauer2021momentum}. 
Note that the experiment probes exciton dynamics, whereas our framework simulates free-carrier scattering and currently omits long-range dipole interactions~\cite{timmer2024ultrafast}. Despite these theoretical simplifications, our model successfully captures the intrinsic depolarization timescale.
Such consistency across different pump chiralities confirms that the valley depolarization lifetime is an intrinsic system property.


\subsection*{Bridging microscopic dynamics to observables via tr-ARPES}
\begin{figure}[!htbp]
  \centering
  \includegraphics[width=0.85\linewidth]{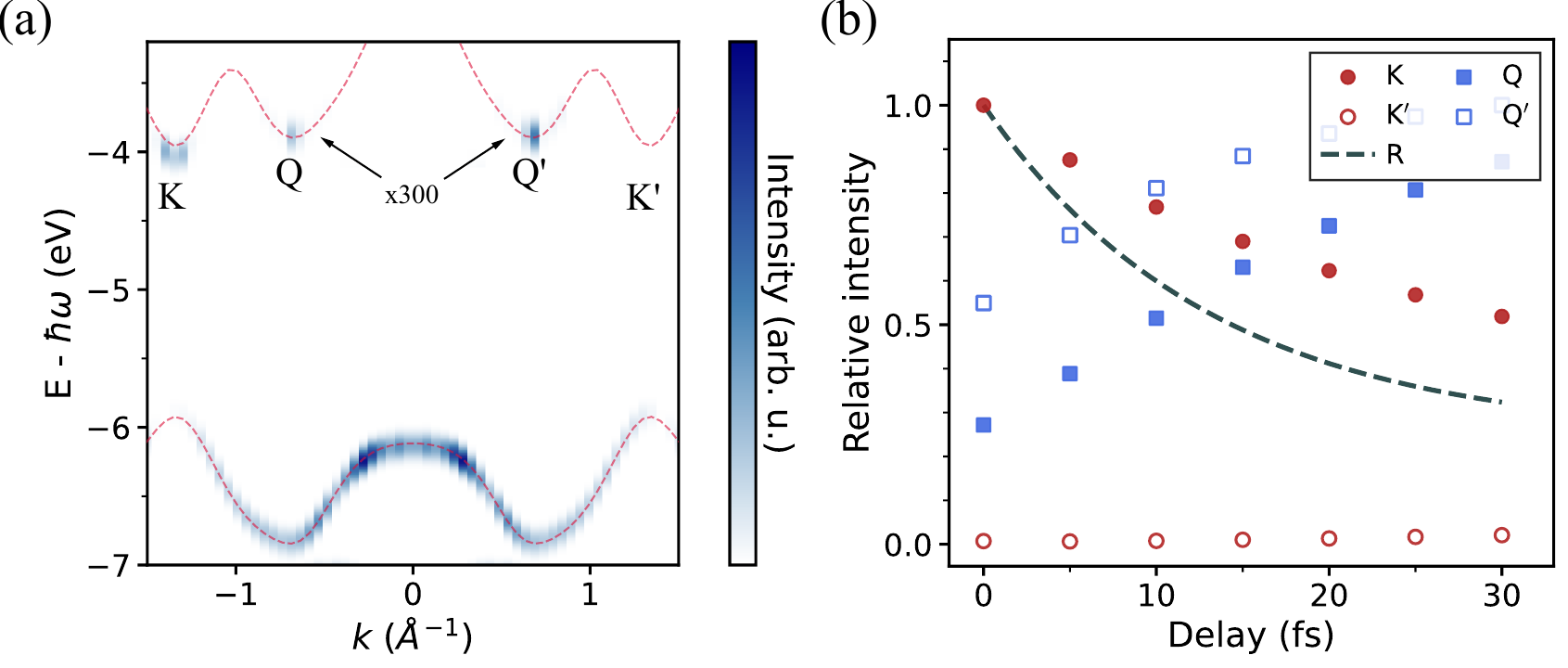}
  \caption{\textbf{ARPES simulations of valley relaxation dynamics.} \textbf{a}, Simulated tr-ARPES spectrum at zero pump-probe delay ($\tau=0$). The spectral intensities at the Q and Q$^\prime$ valleys are scaled by a factor of 300 for clarity. \textbf{b}, Temporal evolution of the relative valley intensities. Solid and open red circles represent the K and K$^\prime$ valleys, respectively, whereas solid and open blue squares denote the Q and Q$^\prime$ valleys. The black dashed line represents the fitted intrinsic population decay.}
  \label{fig:arpes}
\end{figure}

Having resolved the microscopic intervalley scattering pathways, we next ask how these dissipative dynamics would appear in a time-resolved photoemission experiment. This step is essential because the valley populations discussed above are internal theoretical quantities, whereas tr-ARPES provides direct access to the momentum- and energy-resolved evolution of photoexcited carriers. By combining the dissipative rt-TDDFT propagation with the tSURFF photoemission framework~\cite{de2017first,de2015modeling}, we compute fully ab initio tr-ARPES spectra of monolayer WS$_2$, thereby connecting electron-phonon-driven valley relaxation to an experimentally measurable signal.

In the simulation, a 2 eV circularly polarized pump pulse creates the optically excited state, while a 128 eV probe pulse photoemits the excited electrons at a series of pump-probe delays. The delay $\tau$ is defined as the time difference between the onsets of the pump and probe pulses. Because of the computational cost of the photoemission calculation, SOC is neglected in the tr-ARPES simulations.

Figure~\ref{fig:arpes}(a) shows the simulated spectrum at $\tau=0$. The overlaid red dashed line indicates the DFT band structure and agrees well with the photoemission intensity. The $K$ valley is strongly excited, in contrast to the much weaker signal at $K^\prime$, confirming the valley-selective character of the optical excitation. At the same time, finite intensity is also observed at the $Q$ and $Q^\prime$ valleys. These states are not populated by direct optical excitation, but by phonon-mediated intervalley scattering from the initially excited valley. The simulated spectrum therefore resolves the same valley redistribution identified in Fig.~\ref{fig:valleydynamics}, but now at the level of a measurable tr-ARPES observable.

To extract the time-dependent valley dynamics from the photoemission signal, we integrate the ARPES intensity over small momentum-energy windows around the corresponding valley minima. The resulting relative intensities are shown in Fig.~\ref{fig:arpes}(b), after normalization to the initial $K$-valley intensity and to the final $Q^\prime$-valley intensity. The $K$-valley signal decays with delay time, while the signal in the other valleys builds up, closely following the population dynamics of Fig.~\ref{fig:valleydynamics}(c). To account for the finite pump and probe durations, we model the measured intensity $I(\tau)$ as a convolution of the intrinsic population response $R(t)$ with the pump and probe envelopes~\cite{van2004global},
\begin{equation}
    I(\tau) = (I_{\text{pump}} \ast I_{\text{probe}} \ast R)(\tau),
\end{equation}
where $\ast$ denotes convolution. The intrinsic decay is taken as $R(t)=A_1 e^{-t/\tau_0}+A_2$ for $t\ge 0$, with $\tau_0$ the valley depolarization lifetime. The fitted decay, shown as the black dashed line in Fig.~\ref{fig:arpes}(b), yields $\tau_0=13$ fs. This value is shorter than the 20 fs obtained from the SOC-including population dynamics in Fig.~\ref{fig:valleydynamics}(c), consistent with the omission of SOC in the tr-ARPES calculation. Without SOC, additional scattering channels become available, accelerating the apparent valley relaxation.

\section*{Discussion}

The results presented here establish a first-principles route for incorporating electron-phonon dissipation into rt-TDDFT. The method does not replace the coherent real-time Kohn-Sham propagation. Instead, it supplements it with an open-system evolution of the reduced single-particle density matrix, allowing the electronic state to undergo phonon-mediated relaxation and decoherence during the real-time dynamics. In the present implementation, this is achieved through a mixed real-space/active-space representation. The coherent laser-driven dynamics are propagated in real space, while the collision integral is evaluated in an active space where the connection to electron-phonon transition rates remains explicit. This choice reflects the different roles played by the two representations. Real-space propagation is not the most compact basis for electron-phonon scattering, but it provides a flexible framework for nonlinear dynamics, continuum coupling, and time-resolved spectroscopic observables. The active-space density-matrix update makes it possible to introduce microscopic dissipative channels without giving up these capabilities.

The two applications highlight different aspects of this construction. In bulk Si, the method describes the relaxation of a highly non-equilibrium carrier distribution created by optical excitation. The dynamics reveal a separation between rapid momentum redistribution and slower energy relaxation, providing a first-principles picture of hot-carrier cooling driven by electron-phonon scattering. In monolayer WS$_2$, the same framework resolves the phonon-mediated transfer of optically injected carriers between valleys. By combining the dissipative propagation with a tSURFF-based tr-ARPES calculation, these microscopic scattering dynamics are carried through to a directly measurable photoemission signal. The method therefore does not only provide access to internal carrier populations, but connects dissipative nonequilibrium dynamics to experimental observables.

A central feature of the formulation is that population relaxation and coherence decay originate from the same microscopic electron-phonon scattering channels. The diagonal part of the collision integral reduces to the gain--loss structure familiar from Boltzmann transport theory, whereas the off-diagonal part describes the decay of interband coherences. This density-matrix structure is important in ultrafast regimes where coherent excitation, field-driven motion, and scattering occur on comparable time scales. It provides a way to describe dephasing of pump-induced coherences, damping of coherent oscillations, and prethermal carrier dynamics within the same first-principles framework, rather than treating relaxation and decoherence as separate phenomenological ingredients.

Although the present implementation focuses on a thermal phonon bath, the same strategy can be extended to other environmental channels. Coupling to photons, magnons, substrates, or other quasiparticle reservoirs could be incorporated through appropriate collision integrals or self-energy-derived kernels. This would open the way to first-principles simulations of dissipative dynamics in cavity-modified materials, spin-phonon relaxation, phonon-assisted exciton dissociation, and strong-field phenomena where decoherence competes directly with coherent driving. Solid-state high-harmonic generation and pump-probe spectroscopies are particularly natural targets, since their interpretation often depends on relaxation and dephasing mechanisms that are absent in conventional unitary rt-TDDFT.

From a computational perspective, the active-space formulation keeps the additional cost of the dissipative dynamics under control. The coherent propagation retains the real-space and $\mathbf{k}$-point parallelization structure of standard rt-TDDFT, while the collision integral is evaluated only for the subset of states relevant to the nonequilibrium dynamics. The density-matrix update introduces an additional scaling with the number of active states, but this cost is limited by the active-space projection and remains compatible with realistic material simulations. The resulting scheme is therefore a practical compromise: electron-phonon scattering is treated in the representation where energy and momentum transfer are explicit, while the full real-space propagation remains available for nonlinear response and spectroscopic calculations.

The present implementation should be viewed as a controlled realization of a more general open-system rt-TDDFT framework. It relies on the Born--Markov and secular approximations, treats the phonons as a thermal bath, and uses transition rates computed from equilibrium electron-phonon matrix elements. These assumptions define the present scope of the method rather than fundamental limitations. Regimes involving strong non-Markovian effects, hot phonons, coherent nuclear motion, or feedback from evolving ionic coordinates will require extensions of the bath description, the collision kernel, or the coupling between the active-space and real-space representations. Within its current domain of applicability, however, the framework provides a practical first-principles route to phonon-mediated relaxation and decoherence in real-time electronic-structure simulations.

In summary, we have developed a dissipative rt-TDDFT framework that combines coherent real-space Kohn-Sham propagation with active-space electron-phonon collision integrals. The method captures carrier scattering, relaxation, and decoherence from first principles while preserving access to time-resolved observables. By linking microscopic electron-phonon scattering theory to real-time first-principles spectroscopy, this approach provides a practical route for simulating dissipative nonequilibrium dynamics in realistic materials.
\section*{Methodology}
\subsection*{Basis Construction and Hilbert Space Partitioning}

In order to extract physically meaningful observables from the time-propagated TDKS wavefunctions $|\psi(t)\rangle$, we expand the density matrix in the adiabatic basis $\{|\phi_i(t)\rangle\}$, defined as the eigenstates of the instantaneous Hamiltonian $\hat{H}_{\rm KS}(t)$, satisfying $\hat{H}_{\rm KS}(t)|\phi_i(t)\rangle = \varepsilon_i(t)|\phi_i(t)\rangle$. The adiabatic basis yields a physically intuitive picture, which is closely related to the transient band structure. Serving as a co-moving frame, it naturally compensates for the field-driven reversible intraband polarization. Analogous to the Houston state formalism~\cite{houston1940acceleration,Sato2014}, this approach clearly disentangles the non-adiabatic interband carrier excitations from the mere intraband accelerations~\cite{Sato2014}.


Nevertheless, an adiabatic basis is practically incomplete as the TDKS wavefunctions are expanded using a finite number of relevant bands. Truncating the basis set in a finite space inevitably introduces truncation errors, leading to artificial population leakage and unitarity violation.


To resolve this issue while maintaining computational tractability, we rigorously partition the complete single-particle Hilbert space into an active space $\mathcal{P}$ spanned by a finite set of selected adiabatic states $\{|\phi_{i}(t)\rangle\}$, and its complementary residual space $\mathcal{Q}$ spanned by the residual states $\{|d_{\mu}(t)\rangle\}$. The corresponding projection operators are defined as:
\begin{equation}
    \hat{P}(t) = \sum_{i \in \mathcal{P}} |\phi_{i}(t)\rangle \langle\phi_i(t)|, \quad \hat{Q}(t) = \hat{\mathbbm{1}} - \hat{P}(t) = \sum_{\mu \in \mathcal{Q}} |d_{\mu}(t)\rangle \langle d_\mu(t)|,
\end{equation}
where the residual states are explicitly constructed by projecting out the adiabatic components from the individual TDKS wavefunctions:
\begin{equation}
    | d_\mu(t) \rangle = |\psi_\mu(t) \rangle - \sum_{i \in \mathcal{P}} \langle \phi_i(t) | \psi_\mu(t) \rangle |\phi_i(t) \rangle.
\end{equation}
By explicitly constructing the residual states to absorb all out-of-subspace wavefunction components, we strictly preserve the Hilbert space completeness. Consequently, the electronic 1-RDM  can be expanded as a sum of four distinct sector contributions without any truncation loss:
\begin{equation}
    \hat{\rho}^{(1)} = \hat{P}\hat{\rho}^{(1)}\hat{P} + \hat{P}\hat{\rho}^{(1)}\hat{Q} + \hat{Q}\hat{\rho}^{(1)}\hat{P} + \hat{Q}\hat{\rho}^{(1)}\hat{Q} =  \hat{\rho}^{(1)}_{PP} + \hat{\rho}^{(1)}_{PQ} + \hat{\rho}^{(1)}_{QP} + \hat{\rho}^{(1)}_{QQ}.
\end{equation}
We employ the collision integral $\mathcal{I}_{\rm e-ph}(t)$  in the $\hat{\rho}^{(1)}_{PP}, \hat{\rho}^{(1)}_{PQ}$ and $\hat{\rho}^{(1)}_{QP}$ blocks.

\subsection*{Averaged Transition Rate}

In first-principles calculations of electron-phonon scattering, achieving high energy resolution is required to enforce energy and momentum conservation (the physical phonon energies are $\sim$meV). Thus, an ultra-fine Brillouin zone $\mathbf{k}$-grid is needed. However, in rt-TDDFT, propagating the wavefunctions on such a fine mesh is computationally prohibitive. We present here a complementary coarse-graining scheme that enables the mapping of microscopically deterministic electron-phonon scattering rates onto a computationally tractable coarse $\mathbf{k}$-grid.


The microscopic phonon-assisted transition rate from an initial state $(n, \mathbf{k}_\alpha)$ to a final state $(m, \mathbf{k}_\beta)$ is governed by Fermi's Golden Rule:
\begin{equation}
    \begin{aligned}
    W_{(m,\mathbf k_{\beta})\leftarrow (n,\mathbf k_{\alpha})} &= \frac{2\pi}{\hbar }\sum_{\lambda} \left|g^{\lambda}_{mn}(\mathbf k_{\alpha},\mathbf k_{\beta})\right|^2 \Big[n_{\lambda}( \mathbf q)\, \delta\!\left(\varepsilon_{m\mathbf k_{\beta}}-\varepsilon_{n\mathbf k_{\alpha}}-\hbar\omega_{\lambda}(\mathbf q)\right)\delta_{\mathbf k_{\alpha} = \mathbf k_{\beta}-\mathbf q} \\
    &\quad + \big(n_{\lambda}(\mathbf q)+1\big)\, \delta\!\left(\varepsilon_{m\mathbf k_{\beta}}-\varepsilon_{n\mathbf k_{\alpha}}+\hbar\omega_{\lambda}(\mathbf q)\right) \delta_{\mathbf k_{\alpha} = \mathbf k_{\beta}+\mathbf q}\Big].
    \end{aligned}
\end{equation}
Here, $m$ and $n$ are band indices, $\lambda$ denotes the phonon mode, $\mathbf{q}$ is the transferred phonon momentum, and $n_{\lambda}(\mathbf{q})$ represents the equilibrium phonon distribution of the thermal bath. The delta functions strictly enforce energy conservation.

{To render the real-time dissipative dynamics computationally tractable, we evaluate these rates on a fine grid while evolving the density matrix on a coarser grid. To bridge these two distinct scales, we geometrically partition the fine grid into localized momentum patches. Each coarse point $\mathbf{k}$ acts as the representative center of a cluster of neighboring fine points, denoted as $\mathcal{C}(\mathbf{k})=\{\mathbf{k}_{\alpha}\}$. The effective macroscopic transition rate between two coarse-grid points is then straightforwardly defined as the algebraic average of all microscopic fine-grid transitions connecting their respective patches:}

\begin{equation}
    \overline W_{(m,\mathbf k)\leftarrow (n,\mathbf k')}
=
\frac{1}{N_{\mathbf k}N_{\mathbf k'}}
\sum_{\mathbf k_{\beta}\in\mathcal C(\mathbf k)}
\sum_{\mathbf k_{\alpha}\in\mathcal C(\mathbf k')}
W_{(m,\mathbf k_{\beta})\leftarrow (n,\mathbf k_{\alpha})}, 
\label{w_bar}
\end{equation}
where $N_{\mathbf{k}}$ and $N_{\mathbf{k}'}$ are the number of fine points contained within the clusters $\mathcal{C}(\mathbf{k})$ and $\mathcal{C}(\mathbf{k}')$, respectively. In practice, the effective transition rates $\overline{W}$ are evaluated directly by implementing this coarse-graining scheme into a customized version of the EPW\cite{epw} code.

\subsection*{Orbital Continuity and Gauge Transformation}

Spectral decomposition of the time-dependent single-particle density matrix yields the instantaneous eigenvectors and corresponding occupations:
\begin{equation}
    \hat{\rho}^{(1)}(t) |\tilde{\psi}_{i}(t)\rangle = \tilde{f}_{i}(t) |\tilde{\psi}_{i}(t)\rangle.
\end{equation}
However, numerically obtained eigenvectors $\{|\tilde{\psi}_{i}(t)\rangle\}$ are subject to arbitrary phase jumps and non-physical subspace mixing between degenerate or near-degenerate states, which are caused by the numerical eigensolvers. These problems result in a gauge ambiguity, which breaks the temporal continuity of the smoothly evolving wavefunctions, preventing them from being aligned with the reference TDKS wavefunctions.

To restore orbital continuity, we employ a post-diagonalization gauge-fixing procedure based on the orthogonal Procrustes algorithm. We first construct the population-weighted overlap matrix between the newly diagonalized states and the reference TDKS states:
\begin{equation}
    S_{ij} = \sqrt{\tilde{f}_i f_j} \langle \tilde{\psi}_i(t) | \psi_j(t) \rangle.
\end{equation}
By performing a singular value decomposition (SVD) on this overlap matrix, $S = U \Sigma V^{\dagger}$, we extract the optimal unitary transformation matrix, $R = U V^\dagger$. Applying this rotation aligns the states:
\begin{equation}
    \sqrt{f_j^{\rm u}} | \psi_j^{\rm u}(t) \rangle = \sum_i \sqrt{\tilde{f}_i} | \tilde{\psi}_i(t) \rangle R_{ij},
\end{equation}
where $\{|\psi^{\rm u}(t)\rangle\}$ represents the gauge-fixed updated states. This transformation mathematically guarantees the maximum overlap with the original TDKS orbitals, thereby strictly preserving the smooth temporal evolution of the wavefunctions.

\subsection*{Computational Details}

All real-time TDDFT calculations were conducted with the \texttt{Octopus} code. Throughout, we employed HGH norm-conserving pseudopotentials~\cite{vanderbilt1985optimally} and the local density approximation (LDA) exchange-correlation functional~\cite{perdew1981self}.  The simulations were carried out on a real-space grid with a spacing of 0.21~\AA\  and a time step of 2~as. We set the phonon bath temperature to 300~K.

For monolayer WS$_2$, we adopted  the experimental lattice parameter of $a = 3.15~\AA$~\cite{schutte1987crystal}, which well reproduced the relative energy difference between the $K$ and $Q$ valleys. We included spin-orbit coupling explicitly and sampled the Brillouin zone with a $30\times30$ Monkhorst–Pack (MP) $\mathbf{k}$-point grid~\cite{monkhorst1976special}. Circularly polarized and linearly polarized pump pulses with peak intensities of $2.5\times10^{10}$ and $1.0\times10^{11}$~W/cm$^2$, respectively, were employed. Both lasers had an energy bandwidth of 0.5~eV. To probe the transient dynamics, we applied a 128~eV probe pulse with a 0.2~eV bandwidth.

For bulk Si, we used a fully relaxed lattice parameter of 5.38$~\AA$ and an MP $\mathbf{k}$-grid of size 20 $\times$ 20 $\times$ 20. The sample was irradiated by a linearly polarized laser with photon energy 2.5~eV and peak intensity of $2.5\times10^{10}$~W/cm$^2$.

First-principles electron-phonon matrix elements and scattering rates were calculated with  Quantum ESPRESSO (QE)~\cite{giannozzi2017advanced} and EPW~\cite{epw}, using the same crystal structures employed in the rt-TDDFT simulations. A plane-wave kinetic-energy cutoff of 80~Ry was adopted in all calculations.  Phonons for WS$_2$ were first calculated on a $12\times12\times1$ $\mathbf{q}$-point grid and subsequently interpolated by EPW onto very dense $180\times180\times1$ grids of $\mathbf{k}$-points and $\mathbf{q}$-points. For bulk Si, phonons were calculated on a $6\times6\times6$ grid and interpolated onto $40\times40\times40$ $\mathbf{k}$- and $\mathbf{q}$-point grids.

\section*{Data availability}
The input files necessary to reproduce the calculations presented in this work are freely available in the public GitLab repository at \url{https://gitlab.com/zwnie/collision-integral-input}.

\section*{Code Availability}
The customized EPW code used to calculate the averaged transition rates is openly available at \url{https://gitlab.com/supal/epw-octopus-interface}.

\section*{Acknowledgements}
We acknowledge support from the Marie Sklodowska-Curie Doctoral Network TIMES (grant No. 101118915) and SPARKLE736 (grant No. 101169225), and from the Italian Ministry of University and Research (MUR) under the PRIN 2022738 Grant No 2022PX279E 003.

\section*{Author contributions}
State individual contributions.

\section*{Competing interests}
Declare competing interests or state none.



\clearpage


\appendix

\renewcommand{\thefigure}{S\arabic{figure}}
\renewcommand{\thetable}{S\arabic{table}}
\renewcommand{\theequation}{S\arabic{equation}}
\setcounter{figure}{0}
\setcounter{table}{0}
\setcounter{equation}{0}

\clearpage



\setcounter{section}{0}

\setcounter{subsection}{0}

\setcounter{figure}{0}

\setcounter{table}{0}

\setcounter{equation}{0}

\renewcommand{\thesection}{S.\arabic{section}}

\renewcommand{\thesubsection}{S\arabic{section}.\arabic{subsection}}

\renewcommand{\thefigure}{S\arabic{figure}}

\renewcommand{\thetable}{S\arabic{table}}

\renewcommand{\theequation}{S\arabic{equation}}


\renewcommand{\theHsection}{supp.section.\arabic{section}}

\renewcommand{\theHsubsection}{supp.subsection.\arabic{section}.\arabic{subsection}}

\renewcommand{\theHfigure}{supp.figure.\arabic{figure}}

\renewcommand{\theHtable}{supp.table.\arabic{table}}

\renewcommand{\theHequation}{supp.equation.\arabic{equation}}


\begin{titlepage}
    \centering
    \vspace*{1.5cm}
    {\LARGE\bfseries Supplementary Information:\par}
    \vspace{0.5cm}
    {\Large\bfseries
    First-principles electron-phonon scattering in real-time TDDFT
    \par}
    \vspace{1.3cm}
   {\large
Zhengwei Nie$^{1}$\,\orcidlink{0009-0007-1923-2660}\,\contri\corr, Subhojit Pal$^{1}$\,\orcidlink{0009-0003-6356-3409}\,\contri\corr,~
Marti L{\"u}ders$^{2}$\,\orcidlink{0000-0003-4151-4692},~Alexander Buccheri$^{2}$\,\orcidlink{0000-0001-5983-8631},
Hannes H{\"u}bener$^{2}$\,\orcidlink{0000-0003-0105-1427},
Shunsuke A.~Sato$^{3, 2}$\,\orcidlink{0000-0001-9543-2620},
Umberto De Giovannini$^{1, 2}$\,\orcidlink{0000-0002-4899-1304}\,\corr
\par}
\vspace{1cm}
{\small
$^{1}$ Dipartimento di Fisica e Chimica--Emilio Segr\`e,
Universit\`a degli Studi di Palermo,
Via Archirafi 36, 90123 Palermo, Italy
\par}
\vspace{0.4cm}
{\small
$^{2}$ Max Planck Institute for the Structure and Dynamics of Matter,
Luruper Chaussee 149, Hamburg, Germany
\par}
\vspace{0.4cm}
{\small
$^{3}$ Department of Physics, Tohoku University,
Sendai 980-8578, Japan
\par}
{\small
\corr\ Corresponding authors:
\href{mailto:zwnie1008@gmail.com}{zwnie1008@gmail.com},
\href{mailto:palsubhojit429@gmail.com}{palsubhojit429@gmail.com},
and
\href{mailto:umberto.degiovannini@unipa.it}{umberto.degiovannini@unipa.it}
\par}
\vspace{0.5cm}
{\small
\contri\ These authors contributed equally to this work.
\par}
\vspace{1cm}
{\small\today}
\end{titlepage}
\makeatother


\newcommand{\n}{\textbf{n}}
\newcommand{\q}{\textbf{q}}
\newcommand{\p}{\textbf{p}}
\newcommand{\B}{\mathrm{B}}


\tableofcontents

\section{Density matrix propagation with a Bosonic bath}

We consider an electronic system $\rm S$ coupled to a phononic environment $\rm E$ . The total Hamiltonian in the interaction picture is
\begin{equation}
    H_{\text{tot}} = H_{\rm S} + H_{\rm E} + H_{\rm S-E},
    \label{one}
\end{equation}

where \(H_{\rm S}\) describes the non-interacting Kohn Sham electronic system $H_{\rm S}=\sum_{m} \varepsilon_{m} c^{\dagger}_m c_m $, $H_{\rm E} = \sum_{\mathbf q} \varepsilon_{\mathbf q} (b^{\dagger}_{\mathbf q}b_{\mathbf q} + \tfrac{1}{2})$ represents the phonon bath, and the electron-phonon interaction is
\begin{equation}
H_{\rm S-E} = \sum_{\mathbf q}\sum_{m,n} \big( g^{\mathbf q,-}_{mn} c^{\dagger}_m b_{\mathbf q} c_n + g^{\mathbf q,+}_{mn} c^{\dagger}_n b^{\dagger}_{\mathbf q} c_m \big)
\label{two}
\end{equation}
Here \(c_m^{\dagger}(c_m)\) and \(b_{\mathbf q}^{\dagger}(b_{\mathbf q})\) are fermionic and bosonic operators, respectively, and \(g^{\mathbf q,\pm}_{mn}\) denote the electron-phonon coupling matrix elements,
\begin{equation}
    g^{\lambda}_{mn}(k, Q) = \big< mk + Q\big| \Delta_{\q = Q,\lambda} H_{\rm S}\big|nk\big>,\quad g^{\q, +}_{mn} = g^{\p, -}_{nm},\quad \p = (-Q,\lambda) 
    \label{three}
\end{equation}\
where $\q = (Q,\lambda)$ is a combined index labeling $Q$ phonon wave vector, $\lambda$  mode index and $\pm$ corresponds to phonon absorption and emission. In the interaction picture, the dynamics of the full density matrix \(\rho(t)\) obeys the von Neumann (vN) equation,
\begin{equation}
\frac{d}{dt} \rho(t) = -i [H_{\rm S-E}, \rho(t)], \quad \hbar = 1
\label{four}
\end{equation}
The dynamics of the electronic system is obtained by solving  Eq.\,(\ref{four}) and computing $\rho_{\rm  S}$ tracing out bath degree of freedom.  It requires the knowledge of the eigenstates and eigenvalues of the the full electronic system $H_{\rm S}$, which is very cumbersome as we need to solve the many-body Schr\"odinger Equation (SE).

We can instead define the one body reduced density matrix (1-RDM) as $\rho^{(1)}_{mn} = \langle c^{\dagger}_n c_m \rangle = \mathrm{Tr}_{\rm S}[\rho_{\rm S} c^{\dagger}_{n} c_m ]$. Its exact equation of motion derived from the vN equation Eq.\,(\ref{four}) is
\begin{equation}
 \dot{\rho}^{(1)}_{mn} = i\mathrm{Tr}_{\rm S}\mathrm{Tr}_{\rm E}\Big\{\rho(t) [H_{\rm S-E},c^{\dagger}_{n}c_{m}]\Big\} + i\mathrm{Tr}_{\rm S}\Big\{\rho_{\rm S}\Big[H_{\rm S},c^{\dagger}_n c_m\Big]\Big\}
\label{five}
\end{equation}
The above equation is still exact. In general there exist other different scattering mechanisms that influence hot carrier dynamics as energy relaxation and decoherence point of view like, ${\rm  e-e\ (\text{electron-electron}), e- \gamma \, (\text{electron-photon}),\ e-i\,(impurities), \,etc.}$ which preserve a similar structure like Eq.\,(\ref{two}). Onward we denote electron-phonon interaction as $\rm e-ph$.



Under the weak-coupling (Born) approximation, \(\rho(t) \approx \rho_S(t) \otimes \rho_{\text{E}}(t)\). To proceed, we introduce the phonon-assisted density matrix \(\rho^{\mathbf q}_{mn} = \langle c^{\dagger}_n b_{\mathbf q} c_m \rangle\). The equation for \(\dot{\rho}^{(1)}_{mn}|_{\text{e-ph}}\) becomes
\begin{equation}
\dot{\rho}^{(1)}_{mn}|_{\text{e-ph}} = -i \sum_{\mathbf q,\beta} \big( g^{\mathbf q,+}_{\beta m} \rho^{\mathbf q *}_{n\beta} + g^{\mathbf q,-}_{m\beta} \rho^{\mathbf q}_{\beta n} \big) + \text{H.c.}
\label{seven}
\end{equation}
 H.c. denotes the Hermitian conjugate. To solve the above equation, it is essential to understand the dynamics of $\rho^{\q}$, which characterizes the quantum mechanical phase coherence between electrons and phonons. We can write its dynamics as follows:
 \begin{equation}
    \begin{aligned}
        \dot{\rho}^{\q}_{mn} =  -i (\varepsilon_{m} - \varepsilon_{n} + \varepsilon_{\q})\,\rho^{\q}_{mn} + i \sum_{\alpha, \q'}\Bigg\{ g^{\q', -}_{\alpha n} \Big<c^{\dagger}_{\alpha} c_m b_{\q'} b_{\q}\Big> +  g^{\q', +}_{n \alpha} \Big<c^{\dagger}_{\alpha} c_m b^{\dagger}_{\q'} b_{\q}\Big>\\  - g^{\q', -}_{m \alpha} \Big<c^{\dagger}_{n} c_{\alpha} b_{\q} b_{\q'}\Big> - g^{\q', +}_{\alpha m} \Big<c^{\dagger}_{n} c_{\alpha} b_{\q} b^{\dagger}_{\q'}\Big> \Bigg\} - i \sum_{\alpha \gamma} g^{\q', +}_{\alpha \gamma} \Big<c^{\dagger}_{n} c^{\dagger}_{\gamma} c_{\alpha} c_{m}\Big>
    \end{aligned}
    \label{eight}
\end{equation}
This, in turn, involves additional higher-order kinetic variables represented by expectation values of four operators-specifically, electron-phonon two-body reduced density matrices. As a result, an infinite hierarchy (BBGKY) of equations arises.
A correlation expansion is employed to close the hierarchy of equations. We separate the phonon field into coherent and fluctuating parts, \(b_{\mathbf q} = B_{\mathbf q} + \delta b_{\mathbf q}\), where \(B_{\mathbf q} = \langle b_{\mathbf q} \rangle\). Correspondingly,
\begin{equation}
\rho^{\mathbf q}_{mn} = \rho^{(1)}_{mn} B_{\mathbf q} + \delta\rho^{\mathbf q}_{mn},
\label{nine}
\end{equation}
with \(\delta\rho^{\mathbf q}_{mn}\) capturing genuine electron-phonon correlations. Inserting the factorized part of Eq.\,(\ref{nine}) into Eq.\,(\ref{seven}) yields the coherent dynamics
\begin{equation}
    \dot{\rho}^{(1)}_{mn}|_{\rm coh} = -i \sum_{\beta} \big( \varepsilon_{m\beta} \rho^{(1)}_{\beta n} - \rho^{(1)}_{m\beta} \varepsilon_{\beta n} \big)
    \label{ten}
\end{equation}
where $\varepsilon_{ab} = \varepsilon_a \delta_{ab} + \Big(\sum_{\q} g^{\q, -}_{ab} B_{\q} + \text{H.c}\Big)$ is the renormalized single body energy from the coherent phonon field. This renormalization vanishes when $B_{\mathbf q}=0$. For our numerical simulation, we ignore such renormalization.\\


\noindent To account for dissipative processes, we retain the correlation term \(\delta\rho^{\mathbf q}_{mn}\). Its equation of motion involves higher-order correlators, which we factorize using mean-field decoupling. For instance, four-operator averages are approximated as
\begin{equation}
\begin{aligned}    
\Big< c^{\dagger}_{\alpha} c_m b^{\dagger}_{\mathbf q'} b_{\mathbf q} \Big> \simeq & \rho^{(1)}_{m\alpha} \, \delta_{\mathbf q\mathbf q'} \, n_{\mathbf q}(t)\\
\Big<c^{\dagger}_{n} c_{\alpha} b_{\q} b^{\dagger}_{\q'}\Big> \simeq & \rho^{(1)}_{\alpha n}\ \delta_{\q\q'} \big(\n_{\q}(t) + 1\big)\\
\Big< c^{\dagger}_{n} c^{\dagger}_{\gamma} c_{\alpha} c_{m} \Big> \simeq & \rho^{(1)}_{mn} \rho^{(1)}_{\gamma\alpha} - \rho^{(1)}_{m\gamma} \rho^{(1)}_{n\alpha},
\end{aligned}
\label{11}
\end{equation}
assuming no phonon coherence (\(B_{\mathbf q}=0\)) and contribution from the other correlators are zero. This leads to a closed equation for \(\delta\rho^{\mathbf q}_{mn}\):
\begin{equation}
\delta\dot{\rho}^{\mathbf q}_{mn} = -i\sum_{\alpha\gamma} \mathcal{L}^{\mathbf q}_{mn;\alpha\gamma} \, \delta\rho^{\mathbf q}_{\alpha\gamma} + i\mathcal{S}^{\mathbf q}_{mn}(t),
\label{12}
\end{equation}
where
\begin{equation*}
\begin{aligned}
\mathcal{L}^{\q}_{mn;\alpha\gamma}
&= \,\Big(
      \varepsilon_{m\alpha}\,\delta_{n\gamma}
      - \delta_{m\alpha}\,\varepsilon_{\gamma n}
      + \varepsilon_{\q}\,\delta_{m\alpha}\delta_{n\gamma}
    \Big), \\[8pt]
\mathcal{S}^{\q}_{mn}(t)
&= -\,\sum_{\alpha\gamma}
      g^{\q,+}_{\gamma\alpha}\big(\n_{\q}(t)+1\big)\,
      \rho^{(1)}_{\alpha n}(t)\,
      \big(\delta_{m\alpha}-\rho^{(1)}_{m\alpha}(t)\big) \\[4pt]
&\quad +\,\sum_{\alpha\gamma}
      g^{\q,+}_{\gamma\alpha}\,\n_{\q}(t)\,
      \rho^{(1)}_{m\alpha}(t)\,
      \big(\delta_{\gamma n}-\rho^{(1)}_{\gamma n}(t)\big).
\end{aligned}
\end{equation*}
where \(\mathcal{L}^{\mathbf q}\) is a linear operator containing single-particle energy differences and the phonon energy, and \(\mathcal{S}^{\mathbf q}_{mn}(t)\) is a source term depending on \(\rho^{(1)}\) and the phonon occupation \(n_{\mathbf q}(t)\).


Assuming the electron-phonon interaction is switched on at \(t=0\) (\(\delta\rho^{\mathbf q}_{mn}(0)=0\)), Eq.\,(\ref{12}) is formally solved by
\begin{equation}
\delta\rho^{\mathbf q}_{mn}(t) = i\int_0^t d\tau \, e^{-i(\varepsilon_{mn}+\varepsilon_{\mathbf q})(t-\tau)} \, \mathcal{S}^{\mathbf q}_{mn}(\tau).
\label{13}
\end{equation}
where $\varepsilon_{mn} = \varepsilon_m - \varepsilon_n$. Substituting Eq.\,(\ref{13}) into Eq.\,(\ref{seven}) yields a non-Markovian collision integral,
 \begin{equation}
\begin{aligned}
    \dot{\rho}^{(1)}_{mn}\Big|_{\mathrm{incoh}} 
    &= \sum_{l = \pm 1} \sum_{\mathbf{q}} \sum_{a,b,c} 
       g^{\mathbf{q}, +}_{n a} g^{\mathbf{q}, -}_{c b} 
       \int_{0}^{t} dt'
       \Bigg( \n_{\mathbf{q}}(t') + \frac{1}{2} + \frac{l}{2} \Bigg)
       \exp\bigl[-i(\varepsilon_m - \varepsilon_a + l \varepsilon_{\mathbf{q}})(t - t')\bigr] \\
    &\quad \times \bigl(\delta_{m c} - \rho^{(1)}_{m c}(t')\bigr)\rho^{(1)}_{b a}(t') \\
    &\quad - \sum_{l = \pm 1} \sum_{\mathbf{q}} \sum_{a,b,c} 
       g^{\mathbf{q}, -}_{m a} g^{\mathbf{q}, +}_{c b} 
       \int_{0}^{t} dt'
       \Bigg( \n_{\mathbf{q}}(t') + \frac{1}{2} + \frac{l}{2} \Bigg)
       \bigl(\delta_{a b} - \rho^{(1)}_{a b}(t')\bigr)\rho^{(1)}_{c n}(t') \\
    &\quad \times \exp\bigl[-i(\varepsilon_a - \varepsilon_n + l \varepsilon_{\mathbf{q}})(t - t')\bigr] 
    \;+\; \mathrm{H.c.}
\end{aligned}
\label{14}
\end{equation}
This incorporates memory effects, since the time derivative of the density matrix at time ($t$) depends explicitly on its values at all previous times $t'< t$. The direct numerical evaluation of such a memory integral requires storing the full time history of $\rho^{(1)}$, which is computationally expensive. To obtain a time-local kinetic equation, we first rewrite the memory integral appearing in Eq.~(\ref{14}) in terms of relative time, $\tau =  t - t', \quad t'= t - \tau $. Consequently, the first memory integral in Eq.~(\ref{14}) becomes

\begin{equation*}
    \begin{aligned}
        \int_{0}^{t} dt'
       \Bigg( \n_{\mathbf{q}}(t') + \frac{1}{2} + \frac{l}{2} \Bigg)
       \exp\bigl[-i(\varepsilon_m - \varepsilon_a + l \varepsilon_{\mathbf{q}})(t - t')\bigr] \bigl(\delta_{m c} - \rho^{(1)}_{m c}(t')\bigr)\rho^{(1)}_{b a}(t')\\ = \int_{0}^{t} d\tau
       \Bigg( \n_{\mathbf{q}}(t - \tau) + \frac{1}{2} + \frac{l}{2} \Bigg)
       \exp\bigl[-i(\varepsilon_m - \varepsilon_a + l \varepsilon_{\mathbf{q}}) \tau\bigr] \bigl(\delta_{m c} - \rho^{(1)}_{m c}(t -\tau)\bigr)\rho^{(1)}_{b a}(t - \tau)
    \end{aligned}
\end{equation*}
and an analogous expression is obtained for the second integral. Under the Markov approximation, the environmental correlation time \(\tau_{\rm E}\) is assumed to be much shorter than the characteristic evolution of the electronic system $\tau_{\rm S}$, (\(\tau_{\rm E} \ll \tau_{\rm S}\)). Therefore, the density matrix changes only slightly over the memory time, allowing the approximation $\rho^{(1)}(t - \tau) \simeq \rho^{(1)}(t)$ and $n_{\mathbf q}(t-\tau)\simeq n_{\mathbf q}(t).$


The dynamics of phonon $n_{\q} = \big< b^{\dagger}_{\q} b_{\q}\big> - \big< b^{\dagger}_{\q}\big> \big< b_{\q}\big>$ is governed by;
\begin{equation}
    \dot{\n}_{\q} = \Big(i \sum_{m,n} g^{\q, -}_{mn} \rho^{\q}_{nm} +  \text{c.c}\Big) - \Big( i \sum_{m,n} g^{\q, -}_{mn} B_{\q} \rho^{(1)}_{nm} +  \text{c.c}\Big), 
    \label{16}
\end{equation}
Concerning the phonon correlation $\n_{\q}$ in Eq.\,(\ref{16}), its first order dynamics is absent $\dot{\n}_{\q}|_{\rm coh} = 0$ i.e.,
  $\n_{\q}$ is locked to its initial value that is  typically represented by the Bose-Einstein distribution $\n^{\rm eq}_{\q}$.\\
  Hence the slowly varying quantities can be taken outside the time integral,

  \begin{equation*}
      \begin{aligned}
     \int_{0}^{t} d\tau
       \Bigg( \n_{\mathbf{q}}(t - \tau) + \frac{1}{2} + \frac{l}{2} \Bigg)
       \exp\bigl[-i(\varepsilon_m - \varepsilon_a + l \varepsilon_{\mathbf{q}}) \tau\bigr] \bigl(\delta_{m c} - \rho^{(1)}_{m c}(t -\tau)\bigr)\rho^{(1)}_{b a}(t - \tau) \\
       \simeq \Bigg( \n_{\mathbf{q}}^{\rm eq} + \frac{1}{2} + \frac{l}{2} \Bigg) \bigl(\delta_{m c} - \rho^{(1)}_{m c}(t)\bigr)\rho^{(1)}_{b a}(t) \int_{0}^{t} d\tau  \exp\bigl[-i(\varepsilon_m - \varepsilon_a + l \varepsilon_{\mathbf{q}}) \tau\bigr]
      \end{aligned}
  \end{equation*}
  Finally, since the memory kernel decays on the timescale $\tau_{\rm E}$, while the electronic dynamics evolves over a much longer timescale, the upper integration limit can be safely extended to infinity, and adopt a Gaussian correlation function \(\exp[-t^2/(2\tau_{\rm E}^2)]\). As \(\tau_{\rm E} \to \infty\), energy conservation emerges via the \textbf{Sokhotski--Plemelj} theorem:
\begin{equation}
\lim_{\tau_{\rm E}\to\infty} \int_0^{\infty} d\tau \, e^{-i(\varepsilon_{ab}\pm\varepsilon_{\mathbf q})\tau/\hbar} e^{-\tau^2/(2\tau_{\rm E}^2)} = \pi\hbar\delta(\varepsilon_{ab}\pm\varepsilon_{\mathbf q}) - i \hbar\mathcal{P}\!\left(\frac{1}{\varepsilon_{ab}\pm\varepsilon_{\mathbf q}}\right)
\label{15}
\end{equation}
where the real part gives energy-conserving delta functions and the imaginary part contributes to energy renormalization.


\noindent The resulting Markovian equation for the 1-RDM reads
\begin{equation}
\dot{\rho}^{(1)}_{mn} = \dot{\rho}^{(1)}_{mn}|_{\text{coh}} + \dot{\rho}^{(1)}_{mn}|_{\text{incoh}},
\label{17}
\end{equation}
with the incoherent (collision) term given by
\begin{equation}
\dot{\rho}^{(1)}_{mn}|_{\text{incoh}} = \frac{1}{2}\sum_{\alpha m'n'}\sum_{\mathbf q}
\Big[ (\delta_{m\alpha}-\rho^{(1)}_{m\alpha})\,
\mathbb{A}^{\mathbf q}_{\alpha n,m'n'}\,\rho^{(1)}_{m'n'}
- (\delta_{\alpha m'}-\rho^{(1)}_{\alpha m'})\,
\mathbb{A}^{*\mathbf q}_{\alpha m',mn'}\,\rho^{(1)}_{n'n} \Big] + \text{H.c.},
\label{18}
\end{equation}
where the generalized scattering rates are
\begin{equation}
\mathbb{A}^{\mathbf q}_{\alpha n,m'n'} = \frac{2\pi}{\hbar}\sum_{\pm}
\big(n^{\rm eq}_{\mathbf q}+\tfrac12\pm\tfrac12\big)\,
g^{\mathbf q,\pm}_{\alpha m'}\,g^{\mathbf q,\pm *}_{nn'}\,
\delta^{1/2}(\varepsilon_{\alpha m'}\pm\varepsilon_{\mathbf q})\,\delta^{1/2}(\varepsilon_{n n'}\pm\varepsilon_{\mathbf q}).
\end{equation}


Eq.\,(\ref{18}) constitutes a non-linear quantum kinetic equation for the electron density matrix. The scattering rates \(\mathbb{A}^{\mathbf q}\) incorporate both emission (\(n^{\rm eq}_{\mathbf q}+1\)) and absorption (\(n^{\rm eq}_{\mathbf q}\)) processes. Unlike the semiclassical Boltzmann equation, it simultaneously describes the evolution of electronic populations and quantum coherences. In the diagonal limit $\rho^{(1)}_{mn} = f_m \delta_{mn}$, they reduce to the standard Fermi's golden rule. 


\begin{equation}
    \begin{aligned}
        \dot{f}_{m} = \mathcal{I}[f_{m}(t)] = (1 - f_{m}(t)) \sum_{n} \overline{W}_{m \leftarrow n} f_n(t) - f_m(t) \sum_n \overline{W}_{n \leftarrow m} (1-f_n(t))
    \end{aligned}
\end{equation}
where
\begin{equation*}
    \overline{W}_{n \leftarrow m} = \frac{2\pi}{\hbar}\sum_{{\bf q}\lambda,\pm} \big(n^{\rm eq}_{\mathbf q}+\tfrac12\pm\tfrac12\big) |g^{q\lambda}_{mn}|^2 \delta(\varepsilon_m - \varepsilon_n \pm \hbar \omega_{{\bf q}\lambda})
\end{equation*}

For the coherences $m \neq n$, Eq.~(\ref{18}) contains couplings to all other density-matrix elements, $\rho^{(1)}_{m'n'}$ through the generalized scattering rates $\mathbb{A}^{\mathbf q}_{\alpha n,m'n'}$. In the interaction picture these couplings acquire oscillating phase factors of the form $\sim e^{-i (\omega_{mn} - \omega_{m'n'}) t}$, where $\omega_{mn} = (\varepsilon_m - \varepsilon_n )/\hbar$ are the Bohr frequencies of the electronic system. If two coherences have different Bohr frequencies, $\omega_{mn} \neq \omega_{m'n'}$, their mutual coupling oscillates rapidly compared with the relaxation dynamics. Under the well known secular approximation, these fast oscillatory terms are neglected. This approximation is justifies when $|\omega_{mn} - \omega_{m'n'}| \gg \Gamma$, where $\Gamma$ is a characteristic e-ph relaxation or dephasing rate. Equivalently, there must exist a coarse-graining time $\Delta t$ satisfying 
\begin{equation*}
    |\omega_{mn} - \omega_{m'n'}|^{-1} \ll \Delta t \ll \Gamma^{-1}
\end{equation*}
Over such an interval, the rapidly oscillating non-secular terms average to zero, while the slower dissipative evolution remains resolved. This yields the coherences~\cite{stefanucci2013nonequilibrium, haug1994quantum,haug2008quantum,korolev2024unveiling} terms,
\begin{equation}
    \begin{aligned}
        \mathcal{I}[\rho^{(1)}_{mn}(t)] \approx - \frac{\gamma_m + \gamma_n}{2} \rho^{(1)}_{mn}(t)
    \end{aligned}
\end{equation}
where 
\begin{equation*}
\gamma_n = \sum_i \Big[\overline{W}_{i \leftarrow n} (1 - \rho^{(1)}_{ii}(t)) + \overline{W}_{n \leftarrow i} \rho^{(1)}_{ii}(t)\Big]
\end{equation*}

One important point is that Eq.\,(\ref{18}) is a non-linear density matrix equation which is not conventional Lindblad type. To obtain this, we need to consider one further approximation, low-density limit, $(\mathcal{I} - \rho^{(1)}) \approx \mathcal{I}$ in Eq.\,(\ref{18}).

\begin{equation}
    \begin{aligned}
        \dot{\rho}^{(1)}_{mn}\Big|_{\rm e-ph} = \frac{1}{2}\displaystyle\sum_{ m'n'}\sum_{\q} \Bigg[ \mathbb{L}^{\q}_{m n,m'n'} \,\rho^{(1)}_{m'n'} - \mathbb{L}^{\ast\,\q}_{m' m',mn'} \,\rho^{(1)}_{n'n}\Bigg] + {\rm H.C}
    \end{aligned}
    \label{lindblad}
\end{equation}
Both the Eqs.\,(\ref{18},\ref{lindblad}) preserve physically admissible completely positive and trace-preserving~\cite{simoni2025first} (CPTP) dynamics of the density matrix. The Lindblad form is a special case of the more general non-linear equation, which is valid in the low-density limit. The non-linear equation can capture effects beyond the Lindblad approximation, such as population-dependent scattering rates.

\bibliographystyle{naturemag}
\bibliography{bibliography}


\end{document}